\documentclass[11pt]{article}
\usepackage{amssymb}
\usepackage{geometry}                
\usepackage{graphicx}
\usepackage{rawfonts}
\usepackage{amssymb}
\usepackage{epstopdf}
\usepackage{amsmath,amsthm,amssymb}
\usepackage{mathtools}
\usepackage{fullpage}
\usepackage{color}
\usepackage[usenames,dvipsnames,svgnames]{xcolor}
\usepackage[colorlinks=true,citecolor=blue,linkcolor=red,urlcolor=blue]{hyperref}
\usepackage[font=footnotesize, labelfont=bf, margin=0.5cm]{caption}
\usepackage[labelformat=simple]{subcaption}
\usepackage{rawfonts}
\usepackage{etex}
\usepackage{placeins}

\usepackage[normalem]{ulem}

\newtheorem{theorem}{Theorem}

\def\Pr{\noindent \emph{Proof: }}
\def\qed{$\Box$}

\usepackage{graphicx,color}
\usepackage{amsmath,amssymb,latexsym}
\usepackage{mathrsfs}

\def\Ref#1{(\ref{#1})}

\input{prepictex}
\input{pictex}
\input{postpictex}

\begin{document}

\title{Mixing, segregation, and collapse transitions of interacting copolymer rings}

\author{E J Janse van Rensburg$^1$, E Orlandini$^2$, M C Tesi$^3$ and S G Whittington$^4$ \\
$^1$  Department of Mathematics and Statistics, York University,\\ Toronto, Ontario M3J~1P3, Canada \\
$^2$ Dipartimento di Fisica e Astronomia e Sezione INFN, Universit\`a di Padova, \\ 
Via Marzolo 8, I-35131 Padova, Italy \\
$^3$ Dipartimento di Matematica, Universit\`a di Bologna, \\ Piazza di Porta San Donato 5, 
I-40126 Bologna, Italy \\
$^4$ Department of Chemistry, University of Toronto, Toronto M5S 3H6, Canada \\}
\date{\today}

\vspace{10pt}

\maketitle

\begin{abstract}
A system of two self and mutual interacting ring polymers, close together 
in space, can display several competing equilibrium phases and phase 
transitions. Using Monte Carlo simulations and combinatorial arguments 
on a corresponding lattice model, we determine three equilibrium phases, 
two in which the rings segregate in space and are either extended 
(the segregated-expanded phase) or compact (the segregated-collapsed phase). The 
third is a mixed phase where the rings interpenetrate. The corresponding 
phase boundaries are located numerically and their critical nature is discussed.
Finally, by looking at the topological properties of the three phases, we 
show that the two rings are likely to be linked in the mixed phase and 
knotted in the segregated-collapsed phase.
\end{abstract}


\section{Introduction}
\setcounter{equation}{0}

In this paper we investigate a model of two ring polymers that are required 
to be close together in space.  The monomers in the two rings are different 
and this can lead to a mixed phase where the two rings interpenetrate 
\cite{JOTW22}.  In addition, either or both of the rings can collapse to form 
a compact ball.  This model is connected to the behaviour of a diblock copolymer. 

The synthesis of catenated ring polymers
\cite{Ohta2008,Cao2015catenated}, often in 
$AB$-diblock form \cite{niu2009polycatenanes} with one ring
the $A$-block and the other ring the $B$-block, presents a novel situation with 
a phase diagram admitting several possible phases, including segregated, 
collapsed and mixed phases in a dilute solution, in addition to additional 
phases when the diblock catenane is in a melt or in confinement 
\cite{Cao2015catenated}.  The linking of components in a catenane mix 
results in interlocking ring polymer sheets \cite{Luengo2024shape} resembling 
chainmail and held together by mechanically bonded rings due to the 
physical linking of the component ring polymers.  Numerical simulations of 
catenated ring copolymers show that the rings are pulled together by 
catenation, giving rise to smaller radii of gyration (when compared to a 
single ring of the same length) and reduced distance between the 
centres of mass \cite{SC14,JOTW22}.  Furthermore, it has been shown that 
ring-shaped polycatenanes have the capacity to retain a relevant quantity 
of twist, which significantly alters their metric and local 
properties~\cite{tubiana2022circular}.

\begin{figure}[t]
\begin{center}
\includegraphics[width=0.8\textwidth]{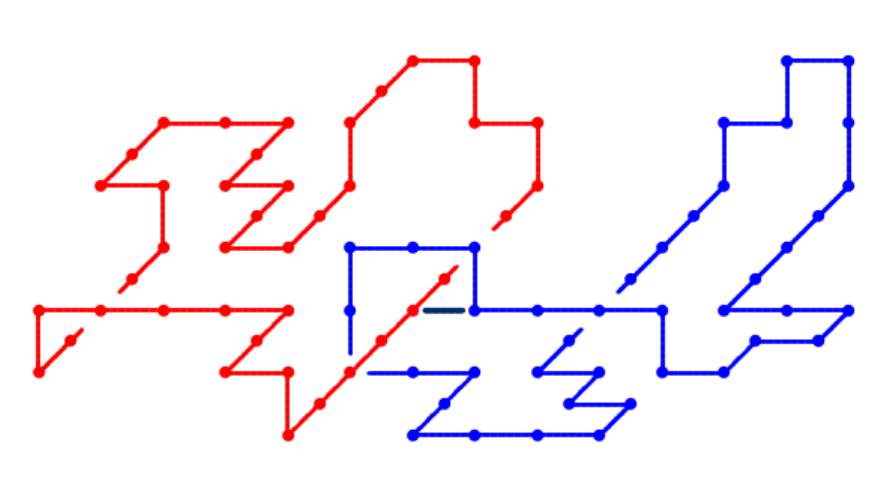}
\end{center}
\caption{\textbf{A cubic lattice model of an $AB$-diblock catenane with an
$A$-block (red) and a $B$-block (blue).  We implement this model by sampling
lattice polygons while keeping the two polygons near each other.  This is
done by having at least one vertex in the $A$-block a unit distance from
a vertex in the $B$-block.  A pair of such vertices are marked with the solid
line in the figure.  The two polygons can be linked (as in this figure) to model a catenane.}}
\label{fig0}
\end{figure}

A cubic lattice model of two ring copolymers in a dilute solution that may be 
catenated consists of two interacting ring polymers (namely, lattice polygons) as
illustrated in figure~\ref{fig0}.  
The two polygons are held in proximity, and we examine metric and linking
properties using Monte Carlo simulations. This approach is broadly similar to the study
in reference~\cite{SC14}, but with additional (repulsive or attractive) short-ranged 
forces between monomers in the two respective component polygons, which are also 
(independently) self-interacting if there are forces between monomers in each.
This distinguishes the two polygons, so that, when catenated, they are a model of
an $AB$-diblock ring polymer, with one ring the $A$-block and the other ring the
$B$-block. 
 
If the self-interacting forces between monomers in each component polygon
are large, then they go through $\theta$-points to collapsed regimes, 
while, if the forces between monomers in each block (one in the $A$-block, 
and the other in the $B$-block) are large, then the two component polygons 
should interpenetrate each other, moving into a mixed regime.  If either of the
$A$- or $B$-blocks are collapsed, then they are in a collapsed or globular
phase with increased self-entanglements as observed by the increased
probability that they are non-trivial knots \cite{tesi1994knotting}.  
In the mixed regime the probability that the components form a catenane 
is increased \cite{orlandini2004entangled,panagiotou2011study,JOTW22}.  
We investigate this by examining the linking probability of the $A$- and 
$B$-blocks as a function of the interaction between the components. 
 
We implement the model in figure~\ref{fig0} by considering two lattice polygons 
(or closed self-avoiding walks) of the same length on the simple cubic lattice.  
The two polygons are held together in the lattice by having at least one 
vertex in one polygon (this is the $A$-block) a unit distance from a 
vertex in the other polygon (namely the $B$-block).  There are three 
energy terms in this model.  If two monomers of type $A$ are unit 
distance apart and not connected by an edge of the polygon, then 
they form an $AA$-\textit{contact} and they contribute an energy 
$\epsilon_{AA}$ to the total energy of the system.  Similarly, two 
monomers of type $B$ a unit distance apart form a $BB$-contact 
and contribute an energy $\epsilon_{BB}$, and a pair of monomers of 
types $A$ and $B$ a unit distance apart form an $AB$-contact and 
contribute an energy $\epsilon_{AB}$.  If $\epsilon_{AA} < 0$ 
this will favour collapse of the $A$-ring into a globular phase, and similarly, 
$\epsilon_{BB} < 0$ will favour collapse of the $B$-ring.  If $\epsilon_{AB} < 0$, 
this will favour intermingling of the two polygons and a transition to 
a mixed phase where we examine the linking probability of the two rings \cite{JOTW22}.

\begin{figure}[t]
\begin{center}
\includegraphics[width=0.8\textwidth,height=0.50\textheight]{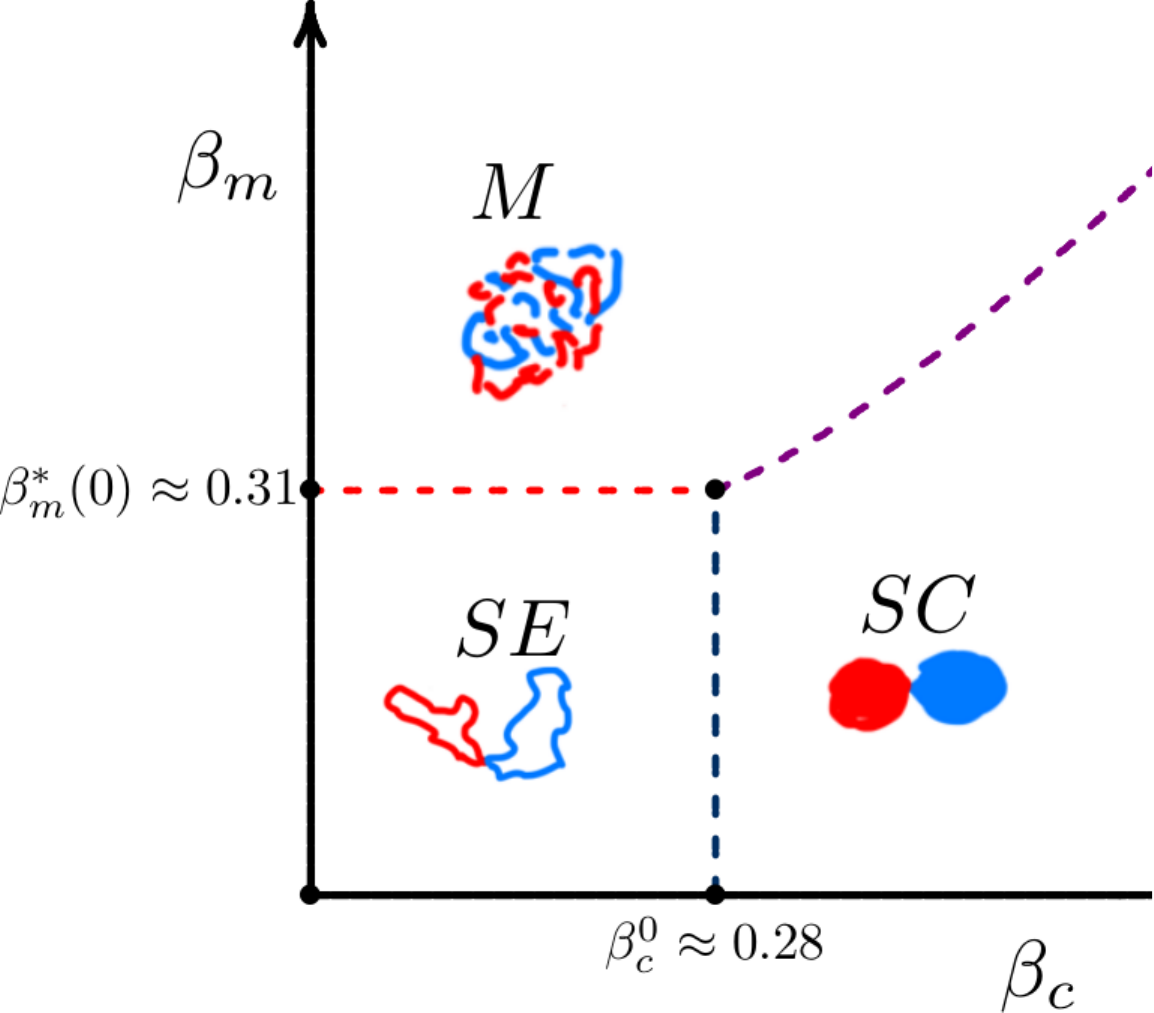}
\end{center}
	\caption{\textbf{A hypothetical sketch of the model's expected phase 
	diagram.  There are at least three possible phases of mixed(M) or 
	expanded(E)-collapsed(C) copolymer.  The segregated-expanded (SE), 
	segregated-collapsed (SC) and mixed (M) phases are explored in this paper.}
	}
\label{fig:sketch_pd}
\end{figure}

If we set $\epsilon_{AA} =\epsilon_{BB}$, then both rings will collapse at the 
same temperature.  In that case we expect at least three phases: A 
segregated-expanded phase where the rings are largely separated in 
space and each is expanded as in a good solvent, a segregated-collapsed 
phase where the two rings are largely separated but each ring has collapsed 
to a compact ball, and a mixed phase where the two rings interpenetrate to 
a significant extent.  The phase diagram is illustrated in figure \ref{fig:sketch_pd}.  
We expect three phase boundaries in the phase diagram.  The phase transition 
between the segregated-expanded and segregated-collapsed phases when 
$\epsilon_{AB}=0$ is relatively well understood.  It is the collapse transition 
for a ring polymer. 

The transition from the segregated-expanded phase to the mixed 
phase was investigated in \cite{JOTW22} but the transition from 
the segregated-collapsed phase to the mixed phase has not 
previously been investigated.  If we write $\beta_m = -\epsilon_{AB}/kT$ 
and $\beta_c = -\epsilon_{AA}/kT= -\epsilon_{BB}/kT$, where 
$k$ is Boltzmann's constant and $T$ is the absolute temperature, 
then we can look at the $(\beta_c, \beta_m)$ plane and the phase 
boundaries will be lines or curves in this plane.

We use several techniques to investigate this phase diagram. 
 In section 3, we use rigorous arguments to establish the existence 
 of a phase boundary as $\beta_m$ is increased, for all $\beta_c < \infty$.  
We write this phase boundary as $\beta_m^*(\beta_c)$ in the phase
diagram.  Note that $\beta_m^*(\beta_c)$ is a function of $\beta_c$
and is the phase boundary separating the mixed $M$-phase from the
segregated $SE$- and $SC$-phases (see figure \ref{fig:sketch_pd}).  Similarly, 
we denote a phase boundary seperating the expanded $SE$-phase and 
the mixed $M$-phase from the collapsed $SC$-phase by $\beta_c^*(\beta_m)$.

If $\beta_c=0$ then the model has a transition from the $SE$-phase
to the $M$-phase at the critical point $\beta_m^*(0)$, where
we understand that $\beta_m^*(0)$ is the critical point for the transition
when $\beta_c=0$.  We similarly define the critical point $\beta_c^0$ 
separating the $SE$- and $SC$-phases when $\beta_m=0$.
It is thought that the transition at $\beta_c^0$ is second order 
and there are good numerical estimates of its 
location \cite{bennett1998exact,tesi1996interacting}.  The estimates are
\begin{equation}
\beta_c^0 = \begin{cases}
0.2782 \pm 0.0070, & \hbox{for collapsing polygons}; \\
0.2779 \pm 0.0041, & \hbox{for collapsing walks}.
\end{cases}
\label{eqn1}
\end{equation}

Our rigorous results leave many questions unanswered and we 
use Monte Carlo methods to obtain additional information.  
There are several quantities that one can examine to probe 
the details of the phase diagram.  These include thermodynamic 
quantities such as the energy and heat capacity, and metric 
properties such as the radii of gyration of the rings (to detect 
collapse), and the mean distance between the centres of 
gravity of the rings (to detect mixing).  In addition, we can 
probe the extent of intermingling or mutual entanglement 
by calculating the linking probability of the two rings.  We 
proceed by sampling and collecting data on the model
along lines crossing the phase boundaries in the phase diagram, 
using a multiple Markov Chain implementation 
\cite{geyer1995annealing,tesi1996monte} of the Metropolis algorithm
\cite{metropolis1953monte} using Verdier-Stockmayer 
\cite{verdier1962monte} and pivot \cite{madras1990monte} 
elementary moves which can change the knot type of lattice polygons 
and linking complexity between lattice polygons.

The plan of the paper is as follows:  In section 2 we give a more detailed 
description of the lattice model that we use.  Section 3 gives some 
rigorous results about the existence of a phase boundary between 
the mixed phase and the two segregated phases, and results about 
the shape of the phase boundary between the two segregated phases.  
We describe the Monte Carlo method that we use in section 
\ref{sec:MonteCarlo} and report our Monte Carlo results about the 
thermodynamic properties and metric properties in section 
\ref{sec:numerical} and about the topological properties in 
section \ref{sec:topology}.  We close with a short discussion in 
section \ref{sec:discussion}.

\section{The model} 
\label{sec:model}

The pair of self-avoiding lattice polygons has total length $2n$, and each 
has length $n$, both embedded in the cubic lattice, $\mathcal Z^3$. 
The polygons are also mutually avoiding and each has a root vertex 
placed such that the two roots are one lattice unit distance apart. 
We label the two polygons A and B and this setup can either describe 
two polymer rings linked together by a single chemical bond or a 
diblock copolymer with a figure eight connectivity. 

Mutual interactions between the two polygons are 
\emph{mutual contacts} (pairs of vertices $(v_A\in A,v_B\in B)$ such that 
the distance between $v_A$ and $v_B$ is one lattice unit).  As above, there is
an interaction energy $\epsilon_{AB}$ for each mutual contact, and the total 
interaction energy from mutual contacts when a given configuration has $k_m$  
mutual contacts is  $E_m = \epsilon_{AB} k_m$.   In addition,  each block is also
subject to an effective self-attraction due to the quality of the solvent.  By 
defining a \emph{contact} within block A as a pair of non-consecutive 
vertices that are a unit distance apart, we have $E_A = \epsilon_{AA} k_A$ 
where  $\epsilon_{AA}<0$ is an effective attractive energy associated with 
each contact within block $A$ and $k_A$ is the number of such contacts. 
Similarly, for block $B$ we have $E_B = \epsilon_{BB} k_B$ with $\epsilon_{BB}<0$.

Define the parameters $\beta_m = -\epsilon_{AB} /kT$, 
$\beta_A = -\epsilon_{AA}/kT$
and $\beta_B = -\epsilon_{BB}/kT$ where $k$ is Boltzmann's constant and 
$T$ the absolute temperature.  Then the partition function of the system is
\begin{equation}
    Z_{2n}(\beta_m,\beta_A,\beta_B) = \sum_{k_m,k_A,k_B} 
    p_{2n}^{(2)} (k_m,k_A,k_B)\,
     e^{\beta_m k_m+\beta_A k_A+\beta_B k_B}.
    \label{full_partition}
\end{equation}
In equation (\ref{full_partition}) the function $p_{2n}^{(2)} (k_m,k_A,k_B)$ denotes 
the number of configurations of the two chemically distinct polygons with total 
length $2n$ having $k_m$ mutual contacts, and $k_A$, $k_B$ (self-)contacts 
within blocks $A$ and $B$ respectively.  This general form accounts for the possibility 
of the two blocks having different affinities with the solvent.  If we restrict 
to the simpler case of $\epsilon_{AA} = \epsilon_{BB} = \epsilon_c$ and define 
$k_c$ to be the total number of contacts, i.e. $k_c=k_A + k_B$, then the 
partition function~\Ref{full_partition} simplifies to 
\begin{equation}
    Z_{2n}(\beta_m,\beta_c) = \sum_{k_m,k_c} 
    p_{2n}^{(2)} (k_m,k_c)\,
     e^{\beta_m k_m+\beta_c k_c},
    \label{partition}
\end{equation}
where $\beta_c = -\epsilon_c/k T$.  When $\beta_m<0$ and $\beta_c<0$ 
entropy should dominate and we do not expect a significant difference from 
the case with $\beta_m=0$, $\beta_c=0$.  Thus, we restrict most of the 
analysis of the equilibrium phase diagram of the system to the 
values $\beta_m\ge 0$ and $\beta_c\ge 0$. In this range, mutual and 
self-attraction compete to determine the conformational properties of the system.

\subsection{Preliminary considerations}

We expect at least three equilibrium phases in the model, determined by
the strength of the self-attraction $\epsilon_c$, relative to the mutual 
attraction $\epsilon_m$:

\begin{description}
    \item[(i)] For $\beta_m$ and $\beta_c$ sufficiently small the system 
    should be in the \emph{Segregated/Extended} (SE) phase, where 
    the two rings are well separated in space and the average number of 
    mutual contacts scaled by the systems size, $\langle k_m \rangle /2n$, 
    should approach zero in the thermodynamic limit ($n\to\infty$). 
    Moreover,  each polygon in the extended phase should display the 
    metric properties of a self-avoiding polygon. Note that, by the 
    pattern theorem \cite{K63}, the self-contacts' density is non-zero,
    although small, in the thermodynamic limit.  Concerning the 
    topological properties,  we expect a negligible amount of mutual 
    (i.e. linking probability and link complexity) and self-topological 
    entanglement (i.e. knotting probability and knot complexity) for the 
    range of ring's lengths considered here ($n\in [48-400]$).
    \item[(ii)] For $\beta_c\gg\beta_m$ and $\beta_m$ still small the 
    diblock copolymer rings should be in the \emph{Segregated/Compact} 
    (SC) phase where the two rings are still well separated in space 
    ($\langle k_m \rangle /2n\to 0$, as $n\to \infty$) but each ring displays 
    a globular conformation (with metric properties of a compact phase). 
    Consequently, the topological self-entanglement should increase 
    dramatically while the mutual entanglements (measured by linking
    of the components) should remain negligible.
    \item[(iii)] Above a critical value $\beta_m^*(\beta_c)$,  the 
    system is expected to be in a   \emph{Mixed} (M) phase where the two rings 
    are significantly intermingled (the density of mutual contacts 
    $\langle k_m\rangle /2n$ is now positive in the thermodynamic limit). 
    The $SE$-$M$ transition occurs at the critical point $\beta_m^*(0) \approx 0.31$ 
    for $\beta_c=0$ ~\cite{JOTW22} but,  for $\beta_c \ne 0 $, we expect 
    $\beta_m^* (\beta_c)$ to be an increasing function of $\beta_c$ for
    large $\beta_c$.
\end{description}

Finally we note that, to some extent, there is a similarity with the 
adsorption and collapse transition problem \cite{VW96,VW98A}. 
In particular, the fact that the density of mutual contacts is negligible below 
the mixing transition could suggest that some features of the boundaries 
between the phases are similar. For instance, we might guess that the 
phase boundary between the SE and the SC phases ($\theta$-point) does 
not depend on the value of $\beta_m$ when $\beta_m < \beta_m^*(0)$. 

A sketch of the possible phase diagram, which we shall study numerically,  
is given in figure~\ref{fig:sketch_pd}. Monte Carlo simulations to carry 
out this investigation involve sampling at several points of the diagram 
sketched in figure~\ref{fig:sketch_pd}, including large positive values 
of $\beta_c$ and $\beta_m$.

\section{Some rigorous results}
\label{section3}

Let $p_n (k_c)$ be the number of lattice polygons of length $n$ steps in the cubic lattice 
with $k_c$ nearest neighbour contacts between vertices.  This defines a 
model of lattice polygons with nearest-neighbour self-interactions and partition function 
\[P_n(\beta_c) = \sum_{k_c} p_n(k_c)\,e^{\beta_c k_c} .\]
It is known that the thermodynamic limit
\begin{equation}
 \lim_{n\to\infty} \frac{1}{n} \log P_n(\beta_c) = \kappa_3(\beta_c) 
\label{2}
\end{equation}
exists in the cubic lattice for all $\beta_c < \infty$. 
More generally, this limit also exists in the hypercubic lattice 
\cite{tesi1996interacting}, in which case the thermodynamic limit is 
denoted $\kappa_d(\beta_c)$.  In the square lattice, it is denoted 
$\kappa_2(\beta_c)$.  Clearly, $\kappa_2(\beta_c) \leq \kappa_3(\beta_c)$.

A dumbbell with a joining edge of length one is formed by placing 
two mutually disjoint lattice polygons and joining them with a 
single edge between a vertex in the first and a vertex in the second.  
If the number of such dumbbells with both polygons of length $n$,
and a total of $k_c$ nearest neighbour contacts within each polygon, 
and $k_m$ nearest neighbour mutual contacts between the pair 
of polygons, is $p_{2n}^{(2)}(k_m,k_c)$, then the partition function is 
\[ Z_{2n}(\beta_m,\beta_c) = \sum_{k_m,k_c} 
p_{2n}^{(2)}(k_m,k_c)\,e^{\beta_mk_m+\beta_ck_c} \]

\begin{theorem}
There exists a critical value $\beta_m^*\geq 0$ (a function of $\beta_c$) 
such that the limit
\[ \lim_{n\to\infty} \frac{1}{2n} \log Z_{2n}(\beta_m,\beta_c) 
= \phi (\beta_m,\beta_c) \]
exists for all $\beta_m < \beta_m^*(\beta_c)$ for all $\beta_c < \infty$. 

Moreover, $\phi(\beta_m,\beta_c) = \kappa_3(\beta_c)$ if 
$\beta_m\leq \beta_m^* (\beta_c)$ and 
$0 \leq \beta_m^* (\beta_c) \leq 2\,\kappa_3(\beta_c) - \kappa_2(\beta_c)$.
\label{thm1}
\end{theorem}

\Pr
A lattice polygon has a lexicographically least vertex (called the bottom 
vertex), and a lexicographically most vertex (called the top vertex). 
Two polygons can be placed and joined into a dumbbell if the bottom 
vertex of the second is one step in the $x$-direction above the top 
vertex of the first.  These conformations correspond to terms with 
$k_m=0$ in the partition function.  Since the partition function of 
each polygon in the dumbbell is $P_n(\beta_c)$, this shows that
\[  P_n^2(\beta_c) \leq Z_{2n}(\beta_m,\beta_c) \]
for all $\beta_m < \infty$ and $\beta_c < \infty$.  By equation \Ref{2},
\[ \kappa_3(\beta_c) \leq \liminf_{n\to\infty} \frac{1}{2n} 
\log Z_{2n}(\beta_m,\beta_c) \]
for all $\beta_m < \infty$ and $\beta_c< \infty$.

To obtain an upper bound we construct a box of side-length $2n$ and 
centre within one step from the centre-of-mass of the first polygon.  
Consider all placements of the second polygon with its centre-of-mass 
within this box.  This gives the bound
\[ (2n)^3 \,P_n^2(\beta_c) \geq Z_{2n}(\beta_m,\beta_c),
\quad\hbox{for all $\beta_m \leq 0$ and $\beta_c < \infty$}. \]
This shows that
\[ \kappa_3(\beta_c) \geq \limsup_{n\to\infty} \frac{1}{2n} 
\log Z_{2n}(\beta_m,\beta_c)\]
for all $\beta_m \leq 0$ and $\beta_c < \infty$.  This shows that the limit
\[ \lim_{n\to\infty} \frac{1}{2n} \log Z_{2n}(\beta_m,\beta_c) 
= \phi (\beta_m,\beta_c) \]
exists for all $\beta_m \leq 0$ and $\beta_c < \infty$ and is equal to 
$\kappa_3(\beta_c)$,  and therefore independent of $\beta_m$, 
when $\beta_m \leq 0$.

We next consider the case that $\beta_m\geq 0$.  Define $q_n(k_c)$ to be the 
number of polygons in the square lattice, of length $n$ and with $k_c$ nearest 
neighbour contacts.  The partition function of this model is
\[ Q_n(\beta_c) = \sum_{k_c} q_n(k_c)\,e^{\beta_c k_c} \]
and the free energy is given by
\[ \kappa_2 (\beta_c) = \lim_{n\to\infty} \frac{1}{n} \log Q_n (\beta_c) \]
and the limit exists for all $\beta_c < \infty$ \cite{van1998collapsing}.

Proceed by placing a square lattice polygon of length $n$ with $k_c$ 
contacts in the $z=0$ plane,  and an exact copy of it, translated 
one step in the $z$-direction in the $z=1$ plane.  Join these into a 
dumbbell by placing an edge between their bottom vertices.  The
number of mutual contacts is now $n{-}1$.  This shows that
\[ q_n(k_c) \leq p_{2n}^{(2)} (n{-}1,2k_c) \]
Multiplying this by $e^{\beta_m(k{-}1) + 2\beta_c k_c}$ and 
summing over $k_c$ gives
\begin{align*}
\sum_{k_c} q_n(k_c)\, e^{\beta_m(n{-}1) + 2\beta_c k_c}
&\leq \sum_{k_c} p_{2n}^{(2)} (n{-}1,2k_c)\, e^{\beta_m(n{-}1) + 2\beta_c k_c} \\
&\leq \sum_{k_m,k_c} p_{2n}^{(2)} (k_m,2k_c)\, e^{\beta_m k_m + 2\beta_c k_c} \\
\end{align*}
Replace $\beta_c$ by $\beta_c/2$, take logarithms, divide by $2n$ and 
let $n\to\infty$ to see that
\[ (\kappa_2(\beta_c) + \beta_m)/2
\leq \liminf_{n\to\infty} \frac{1}{2n} \log Z_{2n} (\beta_m,\beta_c) \]
for all $\beta_m < \infty$ and $\beta_c < \infty$.

If $\beta_m > 2\kappa_3(\beta_c) - \kappa_2(\beta_c) > 0 $ then
$(\kappa_2 (\beta_c) + \beta_m)/2 > \kappa_3(\beta_c)$ showing that
\[  \liminf_{n\to\infty} \frac{1}{2n} \log Z_{2n}(\beta_m,\beta_c) 
> \kappa_3(\beta_c) . \]
Thus, there is a $\beta_m^*(\beta_c) \geq 0$ such that
\begin{align*}
 \lim_{n\to\infty} \frac{1}{2n} \log Z_{2n}(\beta_m,\beta_c) &= \kappa_3(\beta_c),
 \quad\hbox{if $\beta_m \leq \beta_m^*(\beta_c)$ and }  \\
 \liminf_{n\to\infty} \frac{1}{2n} \log Z_{2n}(\beta_m,\beta_c) &> \kappa_3(\beta_c),
 \quad\hbox{if $\beta_m > \beta_m^*(\beta_c)$}  .
\end{align*}
Moreover, $0 \leq \beta_m^* (\beta_c) \leq 2\,\kappa_3(\beta_c) - \kappa_2(\beta_c)$.

This completes the proof. \hfill \qed

\begin{figure}[h!]	
\centering
\includegraphics[width=0.9\textwidth,height=0.27\textheight]{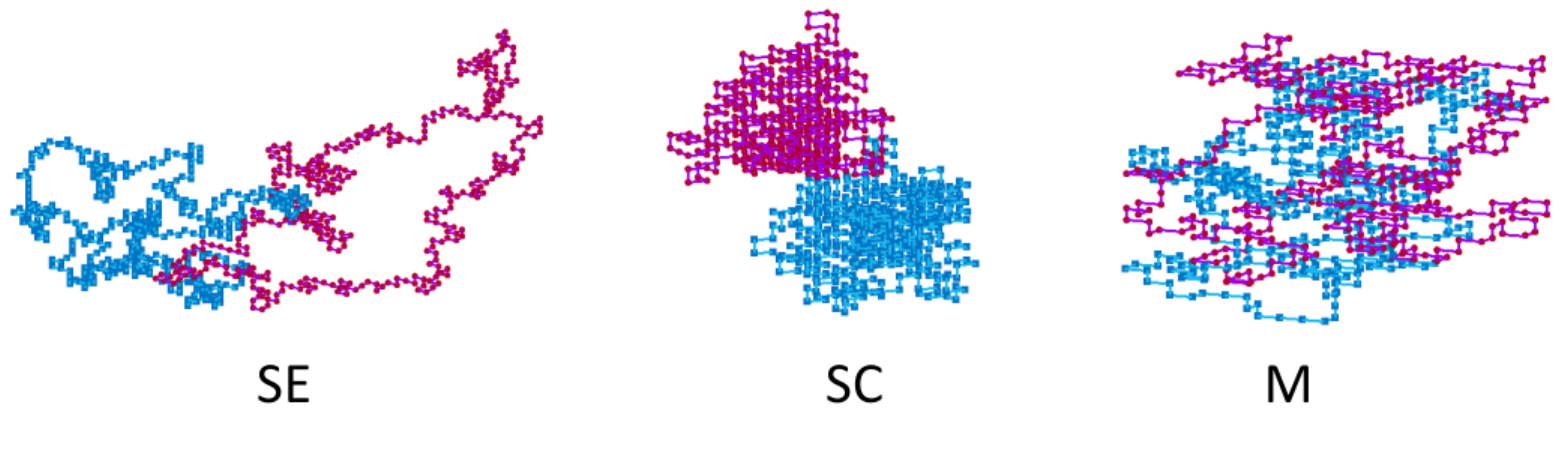}
	\caption{\textbf{Examples of equilibrium configurations of 
	$2n=800$ sampled at $\beta_c=0,\beta_m=0$ (SE), 
	$\beta_c=0.5,\beta_m=0$ (SC) and $\beta_c=0,\beta_m=0.75$ (M).}
	}
\label{fig:confs_examples}
\end{figure}

\bigskip

By Jensen's inequality
\begin{equation}
Z_{2n}(\beta_m,\beta_c)\, Z_{2n}(\beta_m^\prime,\beta_c^\prime)
\geq
\left( Z_{2n}((\beta_m+\beta_m^\prime)/2,(\beta_c+\beta_c^\prime)/2) \right)^2
\end{equation}
Taking logarithms, dividing by $2$, gives
\begin{equation}
(1/2) \log Z_{2n}(\beta_m,\beta_c)
+
(1/2) \log Z_{2n}(\beta_m^\prime,\beta_c^\prime)
\geq
\log Z_{2n}((\beta_m+\beta_m^\prime)/2,(\beta_c+\beta_c^\prime)/2) .
\label{eqn3}
\end{equation}
By the midpoint theorem \cite{hardy1952inequalities} it follows that
$\log Z_{2n}(\beta_m,\beta_c)$ is a convex function of $(\beta_m,\beta_c)$.
Dividing by $2n$ and taking $n\to\infty$ gives the following theorem.

\begin{theorem}
If $\beta_m < \beta^*_m(\beta_c)$ then the free energy 
$\phi(\beta_m,\beta_c)$ is a convex function.  More generally, defining
\[ \limsup_{n\to\infty} \frac{1}{2n} \log Z_{2n}(\beta_m,\beta_c) 
= \phi (\beta_m,\beta_c) \]
is consistent with the definition of $\phi(\beta_m,\beta_c)$ in 
theorem \ref{thm1}, and moreover,  $\phi(\beta_m,\beta_c)$ is 
a convex function for all $(\beta_m,\beta_c)$. \qed
\label{thm2}
\end{theorem}

We notice that the limsup in theorem \ref{thm2} is convex by dividing 
equation \Ref{eqn3} by  $n$ and taking the limsup $n\to\infty$.  Defining
\begin{equation}
\psi(\beta_m,\beta_c) = \liminf_{n\to\infty} \frac{1}{2n} 
\log Z_{2n}(\beta_m,\beta_c)
\end{equation}
observe that $\phi(\beta_m,\beta_c) = \psi(\beta_m,\beta_c)$ for 
all $\beta_m < \beta_m^*(\beta_c)$.  It follows that 
$\psi(\beta_m,\beta_c)$ is monotonic non-decreasing in $\beta_m$ 
and $\beta_c$,  and so it is differentiable almost everywhere.  
Taking the left-limit as $\beta_m \nearrow \beta_m^*(\beta_c)$ shows that
\begin{equation}
\lim_{\beta_m \nearrow \beta_m^*(\beta_c)} \psi(\beta_m,\beta_c)
 = \phi(\beta_m^*(\beta_c),\beta_c) ,
\end{equation}
by the squeeze theorem for limits.  This shows that in the 
statement of theorem \ref{thm2} we can take 
$\beta_m \leq \beta_m^*(\beta_c)$.

We do not know that $\kappa_d(\beta_c)$ is not analytic, rigorously, 
at a collapse critical point $\beta_c^0$.  However, if a collapse transition 
exists at $\beta_c = \beta_c^0$, then $\kappa_3 (\beta_c)$ is non-analytic 
at $\beta_c^0$.  It then follows from Theorem \ref{thm1}
that $\phi(\beta_m,\beta_c)$ is also not analytic at $\beta_c^0$ for all 
$\beta_m \leq \beta_m^*(\beta_c^0)$.   This shows that, if the collapse 
transition exists, then the phase boundary separating the SE and SC 
phases in figure \ref{fig:sketch_pd} must be a straight line.  It is not 
known that the phase boundary separating the SE and M phases is 
a straight line.

\section{Monte Carlo method}
\label{sec:MonteCarlo}
Using a Markov chain Monte Carlo algorithm, we sampled the equilibrium 
conformations of the diblock lattice copolymer model from the Boltzmann 
distribution defined by equation~\Ref{partition}.  The elementary moves
were a combination of pivot moves for self-avoiding polygons that 
ensure ergodicity~\cite{madras1990monte} and local Verdier-Stockmayer 
style moves~\cite{verdier1962monte} that increase the mobility of the 
Markov Chain, especially when the sampling is performed in the regions 
of large positive values of the interaction parameters $\beta_m$ and 
$\beta_c$~\cite{tesi1996monte,tesi1996interacting}.

Sampling was also improved by implementing the elementary moves using 
a Multiple Markov Chain algorithm \cite{geyer1995annealing,tesi1996monte,tesi1996interacting} 
with parallel chains distributed along a sequence of parameters 
$(\beta_c{(j)},\beta_m{(j)})$ for $j=1,2,\ldots,M$. Along each parallel 
chain Metropolis sampling \cite{metropolis1953monte} was implemented to 
sample from the Boltzmann distribution at a fixed pair of values 
$(\beta_c{(j)},\beta_m{(j)})$, and chains were swapped using the 
protocols of Multiple Markov Chain (MCC) 
sampling~\cite{Geyer:1991:CSS,tesi1996monte,tesi1996interacting}.

\begin{figure}[t!]
\centering
\includegraphics[width=0.9\textwidth,height=0.30\textheight]{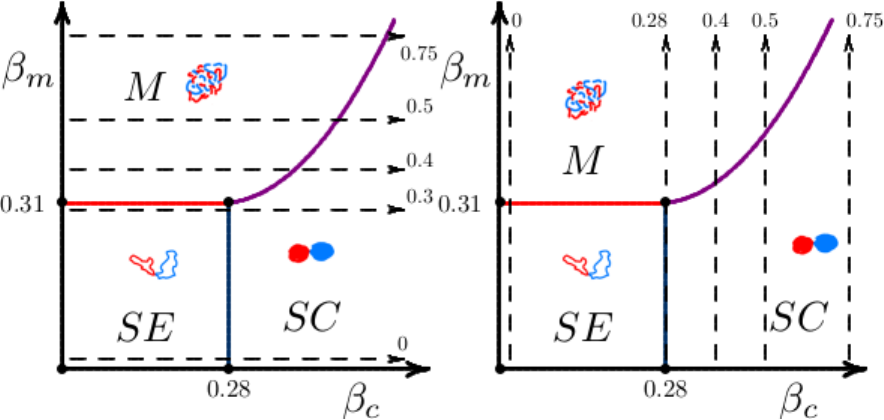}
	\caption{\textbf{Crossing the M-SC boundary by keeping 
	fixed $\beta_m >\beta_m^*(0)$ and varying $\beta_c$ (left) or by 
	keeping fixed $\beta_c>\beta_c^0$ and varying $\beta_m$ (right).}
	}
\label{fig:simulations}
\end{figure}

The stationary distribution of the MCC implementation is the product 
of the Boltzmann distributions along each chain (see 
reference~\cite{tesi1996monte,tesi1996interacting}).  The main difficulty in 
the practical implementation of the MMC scheme is the choice of the sequence
of parameter pairs $(\beta_c{(j)},)\beta_m{(j)}), j=1,2,\ldots,M$.  This is because
the contiguous pairs $(\beta_c{(j)},)\beta_m{(j)})$ and 
$(\beta_c{(j+1)},)\beta_m{(j+1)})$ should be sufficiently close to allow 
a non-negligible number of swaps between them to occur. 
Swaps between chains were proposed at a rate proportional
to the system size ($2n$ -- that is the total combined length of the pair
of polygons, for $n=48, 100, 148, 200, 300, 400$).

We proceeded by collecting data of the metric and thermodynamic properties 
of the model along the chains by either fixing $\beta_m$ and varying $\beta_c$ 
or fixing $\beta_c$ and letting $\beta_m$ vary.  More specifically, we choose a set 
of values of $\beta_m$ at fixed $\beta_c$ or a set of values of $\beta_c$ at fixed 
$\beta_m$ which samples the relevant region of the phase diagram and 
crosses the expected phase boundaries previously described.

For most cases we used $M \approx 25$ parallel chains, and we were 
able to obtain sufficiently uncorrelated samples for systems of 
\textit{total} size $2n\in \{96,200,296,400,600,800\}$. We counted 
a single iteration as $O(1)$ pivot moves, and $O(n)$ local Verdier-Stockmayer 
style moves.  By spacing the reading of data along each Markov chain, 
we were able to sample approximately $2.5\times 10^4$ conformations 
that are essentially \textit{uncorrelated} at each fixed value of the 
pair $(\beta_c^{(j)},\beta_m^{(j)})$ for a total of at least 
$1.25 \times 10^6$ data points for each value of $n$.  The simulations 
were expensive in terms of CPU time.    For example, for $2n=800$
(each polygon in the pair has length $400$)
and $M\approx 25$, a total of $25,\!000$ uncorrelated conformations 
were sampled over two months of CPU time on an $x86\_64$-based 
workstation.  In figure~\ref{fig:confs_examples} we show some 
examples of conformations taken along the chains for different 
values of the pair $(\beta_c^{(j)},\beta_m^{(j)})$.

To explore the phase diagram we primarily carried out simulations 
as shown in figure~\ref{fig:simulations}.  In the first instance 
we fixed $\beta_m$ at values in $\{0,0.2,0.3,0.4,0.5,0.75\}$ and 
collected data along Markov chains for $\beta_c\in [0,1]$.  This is shown 
by the arrows crossing the SE-SC and M-SE phase boundaries in 
the left panel of figure~\ref{fig:simulations}.  For example, the SE-SC 
phase boundary was explored fixing $\beta_m$ at values $0$, $0.2$ 
and $0.3$.  In the right panel we fixed $\beta_c$ instead at the values 
$\{0,0.2,0.28,0.4,0.5,0.75\}$ respectively, while varying $\beta_m$ 
to cross the SE-M and SC-M phase boundaries.

\section{Numerical results}
\label{sec:numerical}

In this section we explore in more detail the phase diagram illustrated 
in figure \ref{fig:sketch_pd} using Monte Carlo simulations.  We will in 
particular focus on the phase boundaries and the thermodynamic
properties of the three phases.

\begin{figure}[h!]	
\includegraphics[width=0.95\textwidth,height=0.825\textheight]{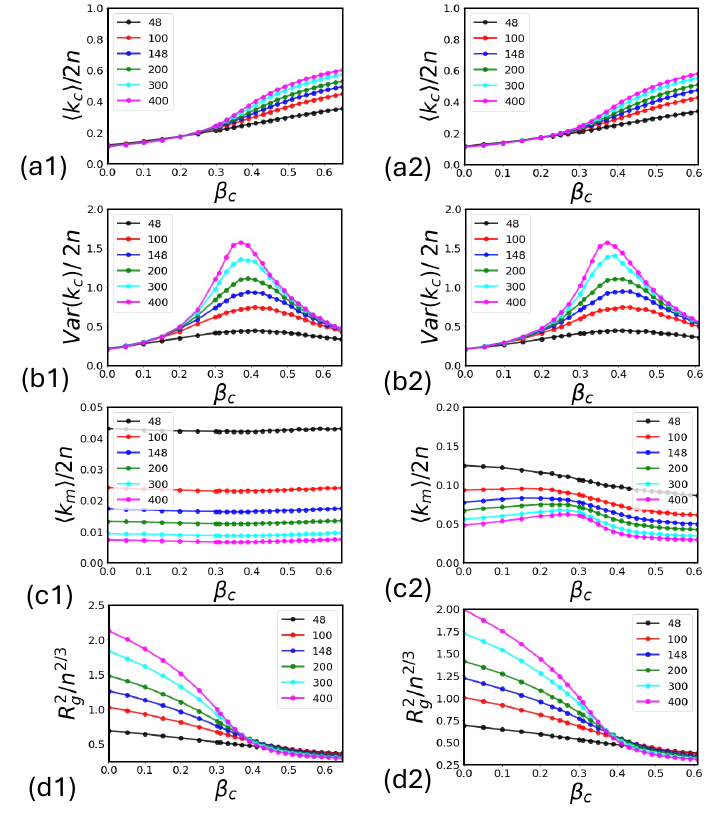}
    \caption{\textbf{(a1-a2) The average number of self contacts scaled by 
    length, $\langle k_c \rangle / 2n$, as a function of $\beta_c$. Different 
    symbols colours correspond to different system sizes (see the legends).    
    (b1-b2) Plots of the corresponding variance of $k_c$ scaled by 
    the system's size, $Var(k_c)/2n$, as a function of $\beta_c$. 
    (c1-c2) Corresponding plots of the average number of 
    mutual contacts scaled by $2n$, $\langle k_m \rangle / 2n$, as 
    a function of $\beta_c$.
    (d1-d2) Plots of the mean-squared radius of gyration of the 
    single components, $\langle R_g^2\rangle$, scaled by $n^{2/3}$, as a 
    function of $\beta_c$.  The two columns correspond to simulations 
    performed at the fixed values of $\beta_m=0$ and $0.30$ respectively.}
    }
\label{fig:SE_SC_thermo}
\end{figure}

\subsection{The SE-SC phase boundary}

Consider first the case that $\beta_m=0$.  
In this event the only nearest-neighbour self-attraction between monomers 
is when $\beta_c > 0$.   In the event that $\beta_c \leq 0$ the model is in 
a phase of segregated and expanded polygons.  Increasing $\beta_c>0$ 
the model is expected to undergo a collapse transition at the $\theta$-point 
at a critical value $\beta_c^0 >0$.  In the case of cubic lattice polygons 
$\beta_c^0\approx 0.278$~\cite{tesi1996interacting,bennett1998exact}, and 
one expects this to be also the location of the $\theta$-point in our model 
when $\beta_m\leq 0$.

In the top row of graphs in figure~\ref{fig:SE_SC_thermo} the average 
number of self-contacts per unit length $\langle k_c \rangle/2n$, for polygons 
of length $n$, is plotted as a function of $\beta_c\in [0,0.65]$.  The values of 
$n$ increase from $n=48$ to $n=400$, and the panels correspond to fixed 
values of $\beta_m=0$ (panel a1) and $\beta_m=0.3$ (panel a2).  The 
data in the two panels consistently show linear growth of $\langle k_c \rangle$ 
for $\beta_c\lesssim 0.3$ and superlinear growth when $\beta_c\gtrsim 0.3$.

If $\beta_m=0$ then the model is composed of two polygons modelling ring 
polymers in a good solvent.  It can be shown that the connective constant 
of the pair of polygons is equal to that of a single polygon.  In the case of a 
single polygon the pattern theorem \cite{K63} shows that $\langle k_c \rangle$ 
grows at least linearly with $n$.  The linear growth seen in 
figure~\ref{fig:SE_SC_thermo}(a1) is then consistent with an expanded 
phase in the model.  The superlinear growth in $n$ for large $\beta_c$ indicates 
nearest neighbour contacts between vertices far apart along the polygon 
backbones -- this is consistent with compact conformations.  This change 
in behaviour at a critical point $\beta_c^0$ is consistent with a $\theta$-point
transition and the locations at which the curves deviate from linear behaviour 
as finite-size estimates of the phase boundary.  The variances 
$Var(k_c) = \langle k_c^2\rangle - \langle k_c \rangle ^2$  for the same 
fixed values of $\beta_m$ are shown in figure~\ref{fig:SE_SC_thermo}(b1-b2).  
As $n$ increases the locations of the peaks shift towards lower values of 
$\beta_c$ while their height increases.  The peaks are finite size effects 
observed due to a singularity in the thermodynamic limit.

The panels in figure~\ref{fig:SE_SC_thermo}(c1-c2) show the average number 
of mutual contacts per unit length, denoted $\langle k_m\rangle/2n$.  This 
decreases with increasing $n$ and may approach zero as $n\to\infty$, indicating
that the model is in a segregated phase for the entire range of $\beta_c$ 
in the simulation.  

In the segregated phases one may distinguish between the extended and 
the compact phases by examining the scaling behaviour of the mean 
squared radius of gyration $\langle R_g^2 \rangle$.  In the extended 
phase  it is known that $\langle R_g^2 \rangle \sim n^{2\nu}$ with 
$\nu = 0.587...~\cite{Clisby:2010:Phys-Rev-Lett}$.  In the compact 
phase one expects that $\langle R_g^2 \rangle \sim n^{2/3}$.  
That is, the curves $\langle R_g^2 \rangle/n^{2/3}$ plotted against 
$\beta_c$ should increase with $n$ for $\beta_c<\beta_c^0$, but inside 
the compact phase they should be independent on $n$.  This behaviour 
is confirmed in figures~\ref{fig:SE_SC_thermo}(d1-d2).  In the compact phase 
the curves do not fully collapse onto a unique curve but show a slight 
decrease with $n$. Nevertheless, the curves cross one another and 
we consider the location of these crossings as a further finite-size estimate of 
the SE-SC phase boundary.

\begin{figure}[t!]	
\includegraphics[width=1\textwidth]{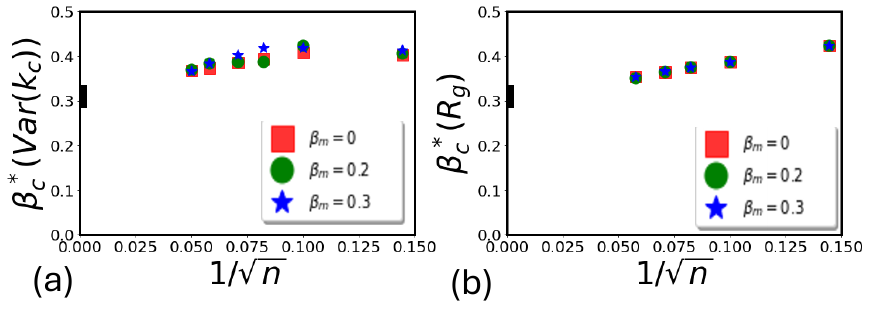}
	\caption{\textbf{Estimating the location of the critical point $\beta_c^0$
	from the locations of peaks in $Var(k_c)$ (denoted by $\beta_c^*(Var(k_c))$)
	and the locations of crossing points of mean square radius of gyration curves
	(denoted by $\beta_c^*(R_g)$) -- see figure \ref{fig:SE_SC_thermo}
       panels (b1) and (b2), and panels (d1) and (d2).}
       \textbf{(a) The locations of the peaks of $Var(k_c)/2n$ for three 
	values of the mutual attraction parameter $\beta_m = 0.0,0.20,0.30$. 
	(b) The corresponding locations of the crossings between the curve 
	$\langle R_g^2\rangle/n^{2/3}$ vs $\beta_c$ at $n=400$ and the curves 
	for $n=48,100,148,200$ and $300$. The three values of $\beta_m$ are 
	the same as in (a). In both cases the data are plotted against $1/\sqrt{n}$. 
	The vertical bar at $1/\sqrt{n}=0$ refers to the estimate of $\beta_c^0$ 
	reported for the standard $\theta$-point in~\cite{tesi1996interacting}.}
	} 
\label{fig:betac_estimates_SE_SC}
\end{figure}

In figure~\ref{fig:betac_estimates_SE_SC}(a) we plot the estimated 
locations of the maxima in the variances $Var(k_c)/2n$.  These maxima were
estimated using a fourth order spline fit to the data points.  We plot
these maxima against $1/\sqrt{n}$ (at the $\theta$-point the finite size 
crossover exponent $\phi$ is expected to have its mean field value $1/2$).
The crossings between the scaled mean square radius of gyration curves 
$\langle R_g^2\rangle/n^{2/3}$ for $n=400$ with the corresponding curves 
estimated for $n=48,100,148,200$ and $300$ are plotted in 
figure~\ref{fig:betac_estimates_SE_SC}(b). 
The thick segment at $1/\sqrt{n}=0$ in the two graphs marks the 
estimated location of the cubic lattice $\theta$-point 
$\beta_c^0 = 0.2782 \pm 0.007$ in reference \cite{tesi1996interacting}.  
Observe that (i) the $n\to\infty$ extrapolations of the data in both
graphs are very close to previous estimates of the cubic lattice 
$\theta$ transition; and (ii) the phase boundary between the SE and SC 
phases seems to be independent of the strength of the mutual 
attraction $\beta_m$, at least in the range $\beta_m\in [0,\beta_m^*(0)] 
\approx [0,0.31]$.

\subsection{SE-M phase boundary}
\label{SE-M}

\begin{figure}[t!]	
\includegraphics[width=1\textwidth]{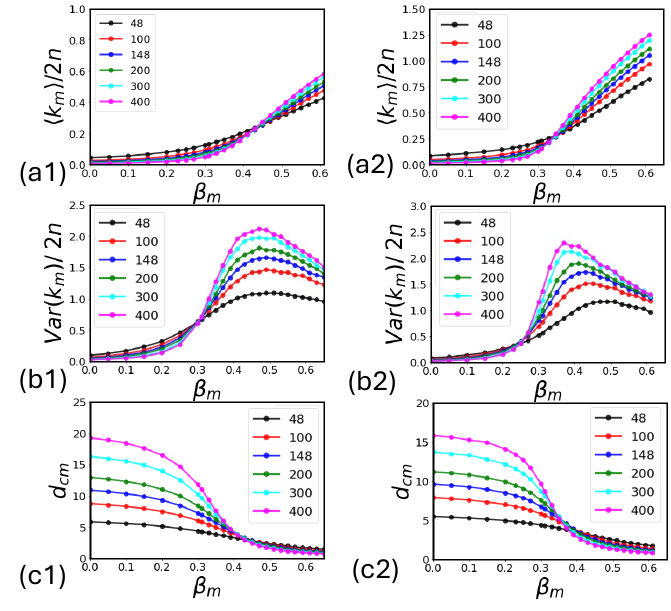}
	\caption{ \textbf{(a1-a2) Plots of the average number of mutual 
	contacts scaled by $2n$, $\langle k_m \rangle / 2n$, as a function 
	of $\beta_m$. Different symbols refer to different system sizes (see legend).    
	(b1-b2) Plots of the corresponding variance of $k_m$ scaled by 
	the system's size, $Var(k_m)/2n$, as a function of $\beta_m$.  
	(c1-c2) Corresponding plots of the distance between the centre of 
	mass of the two polygons $d_{cm}$ as a function of $\beta_m$. 
	The two columns correspond to simulations performed at the fixed values 
	of $\beta_c=0.0$ and $0.28$ respectively.}
	}
\label{fig:SE_ME_thermo}
\end{figure}

If $\beta_c=0$ then the model reduces to two polygons with a mutual 
nearest neighbour attraction between vertices in different polygons.  
In this case the model goes through a segregated-mixed phase transition 
at $\beta_m^*(0)\approx 0.31$~\cite{JOTW22}. 
The average number of mutual contacts per unit length, 
$\langle k_m \rangle / 2n$, is shown as a function of $\beta_m$ 
(for $\beta_c=0$ and $0.28$) in figure~\ref{fig:SE_ME_thermo} 
for $n=48,100,148,200,300$ and $400$.

The first column corresponds to $\beta_c=0$
and the second column to $\beta_c=0.28$. The second row in 
figure~\ref{fig:SE_ME_thermo} corresponds to the
variance of $k_m$ per unit length ($Var(k_m)/2n 
= (\langle k_m^2\rangle - \langle k_m\rangle^2)/2n$) and the last row is
the distance $d_m$ between the centres of mass of the two polygons.

The intersecting variance curves in the second row of graphs are consistent
with a critical point corresponding to a transition into a mixed
phase.  The intersections are also consistent with the estimate 
$\beta^*_m (\beta_c) \approx 0.31$ for $\beta_c\in[0,\beta_c^0]$.

If $\beta_m<\beta_m^*(0)$ then $\langle k_m \rangle / 2n$ 
tends to zero as $n$ increases (this is seen in the left-most panel in the
first row in figure~\ref{fig:SE_ME_thermo}). This behaviour persists 
also for $0< \beta_c  < \beta_c^0$ (as seen in the right-most
panel in the top row of figure~\ref{fig:SE_ME_thermo}). These results are 
consistent with two polygons being segregated in space if $\beta_m < \beta_m^*(0)$. 
Notice that above a particular value of $\beta_m$ the curves 
$\langle k_m \rangle / 2n$ increase with $n$ in the panels in the top row of
figure~\ref{fig:SE_ME_thermo}.  This increase in mutual contacts characterizes
a mixed phase where the two components share the same region in space
(when the model transitions as $\beta_m$ increases beyond the critical point
$\beta_m^*(0)\approx 0.31$ in figure~\ref{fig:sketch_pd} separating the 
SE- and M-phases).

\begin{figure}[t!]	
\includegraphics[width=1\textwidth]{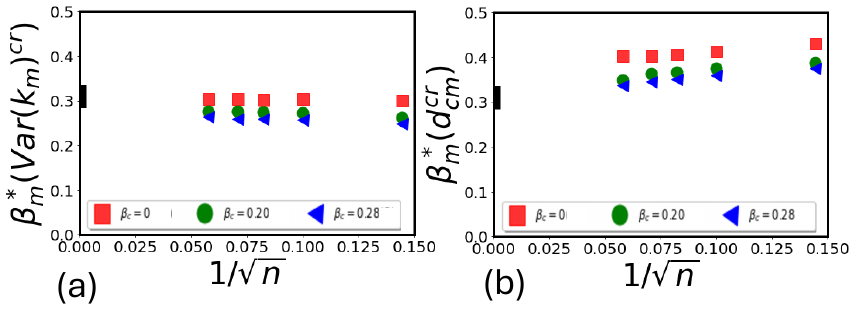}
	\caption{\textbf{Determining the location of the critical point $\beta_m^*(0)$
	from the locations of crossings between the $Var(k_m)$ curves 
	(denoted by $\beta_m^*(Var(k_m)^{cr})$) and from the locations of 
	crossing points of $d_{cm}$ curves (denoted by $\beta_m^*(d_{cm}^{cr})$)
	-- see figure \ref{fig:SE_ME_thermo} panels (b1) and (b2), and panels (c1) and (c2).
	(a) The locations of the crossings of the variance per unit length
	curves $Var(k_m)/2n$ as a function of $n=400$ with the corresponding curves 
	for $m=48,148,200,300$, 
	\bf (b) As in the left panel but for the curves $d_{cm}$ against $\beta_m$. 
	The three values of $\beta_c$ are $0.0,0.20$ and $0.28$ (see legend). 
	In both panels data have been plotted against $1/\sqrt{n}$. The vertical bar 
	at $1/\sqrt{n}=0$ is the estimate of $\beta_m^*(0)$ 
	(see reference~\cite{JOTW22}) for the $\beta_c=0$ case.
       These data appear to approach the critical point $\beta_m^*(0)$ as
	$n$ increases.}
	} 
\label{fig:betam_estimates_SE_ME}
\end{figure}

\begin{figure}[t!]	
\includegraphics[width=1\textwidth]{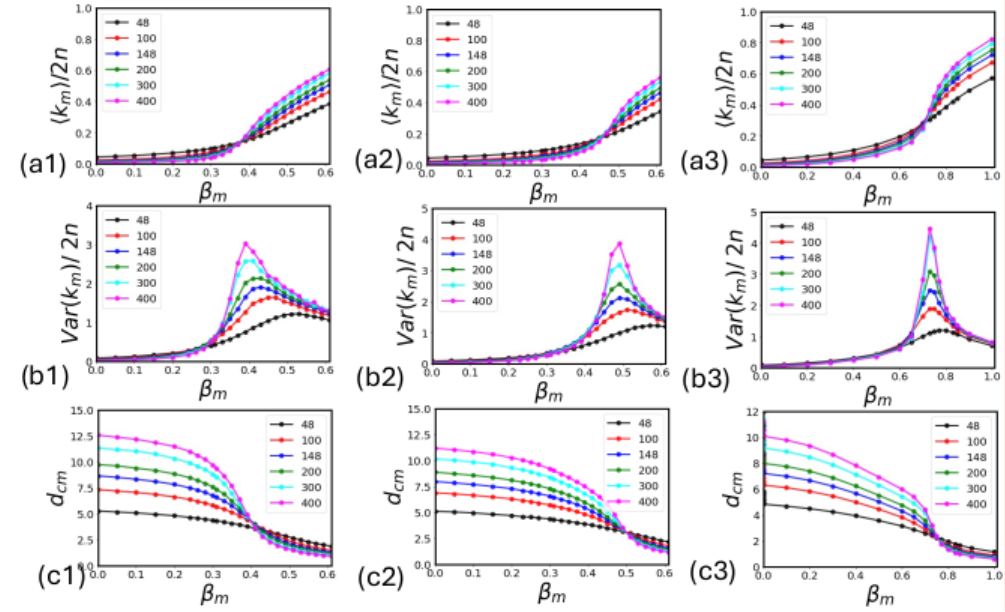}
	\caption{ \textbf{(a1-a3) Plots of the average number of mutual 
	contacts per unit length, $\langle k_m \rangle / 2n$, as a function 
	of $\beta_m$ for  $\beta_c=0.4,0.5$ and $0.75$. These values are 
	well above the theta transition $\beta_c^0\approx 0.28$ (see the right 
	panel of figure~\ref{fig:simulations}). Different symbols refer to different 
	system sizes (see the legend).    
	(b1-b3) Plots of the variance per unit length, $Var(k_m)/2n$, 
	as a function of $\beta_m$ for the same fixed values $\beta_c=0.4$, $0.5$
	and $0.75$.
	(c1-c3) Corresponding $\beta_m$ dependence of the distance 
	between the centre of mass of the two polygons, $d_{cm}$.}
	}
\label{fig:CS_CM_thermo}
\end{figure}

\begin{figure}[t!]	
\includegraphics[width=1\textwidth]{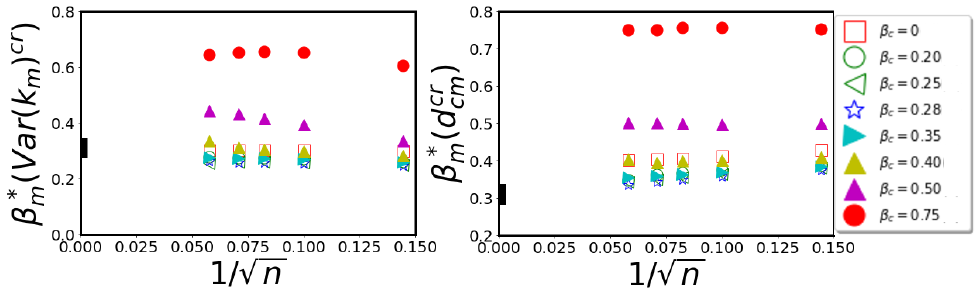}
	\caption{\textbf{Determining the locations of points on the critical curve
	$\beta_m^* (\beta_c)$ by extrapolating the locations of crossings between 
	the $Var(k_m)$ curves (denoted by $\beta_m^*(Var(k_m)^{cr})$) and 
	extrapolating the locations of crossing points of $d_{cm}$ curves (denoted by 
	$\beta_m^*(d_{cm}^{cr})$) -- see figure \ref{fig:SE_ME_thermo} 
	panels (b1) and (b2), and panels (c1) and (c2).
	and figure \ref{fig:CS_CM_thermo} panels (b1), (b2) and (b3), 
	and panels (c1), (c2) and (c3).}
	\textbf{Left panel: The locations of crossings 
	$\beta_m^*(Var(k_m)^{cr})$
	between the curve $Var(k_m)/2n$ for $n=400$ (see the middle row of 
	figure~\ref{fig:CS_CM_thermo}) and $Var(k_m)/2n$ for 
	$n\in\{48,100,148,200,300\}$ for values of $\beta_c < \beta_c^0$ (below the 
	$\theta$-point; these are the empty symbols) with those above 
	$\beta_c^0$ (solid symbols).  The locations are plotted as a function
	of $1/\sqrt{n}$.  Notice that the data below the 
	$\theta$-point tends to line up and extrapolate as $n\to\infty$ to the marked
	solid vertical bar at $1/\sqrt{n}=0$ (this is the estimate for $\beta_m^*$
	in reference~\cite{JOTW22} for the case $\beta_c=0$.  
	In contrast, the data above the $\theta$-point extrapolate to 
	larger values on the vertical axis.  
	Right panel:  This is similar to the left panel, but now for the curves
	$d_{cm}$ in the middle row of  figure~\ref{fig:CS_CM_thermo}.
	The values of $\beta_c$ are reported in the legend.} 
	}
\label{fig:CS_CM_betac}
\end{figure}

The onset of a thermodynamic phase transition between the SE- and M-phases
is also suggested by the $\beta_m$ dependence of $Var(k_m)/2n$ in the curves
in the middle row of panels in figure~\ref{fig:SE_ME_thermo}.  Notice 
the onset of peaks whose heights increase with $n$.   Moreover, the curves 
cross one another in a narrow region of the mutual interaction 
parameter $\beta_m$.  This is consistent with the theorem that,
for $\beta_c<\infty$, and $\beta_m<\beta_m^*(0)$, $\lim_{n\to\infty} Var(k_m)/2n
=0$; see section \ref{section3} and theorem \ref{thm1}.  That result shows 
the presence of an asymmetric phase transition when $\beta_c<\infty$ somewhat 
similar to the adsorption transition of a self-avoiding walk anchored to an 
impenetrable attractive wall \cite{HTW82}. In these cases it is known that 
the intersections of the curves $Var(k_m)/2n$ plotted against $\beta_m$ 
provide a faithful finite-size estimate of the location of the 
phase boundary~\cite{buks_book_2015}.

\begin{figure}[t!]	
\includegraphics[width=1\textwidth]{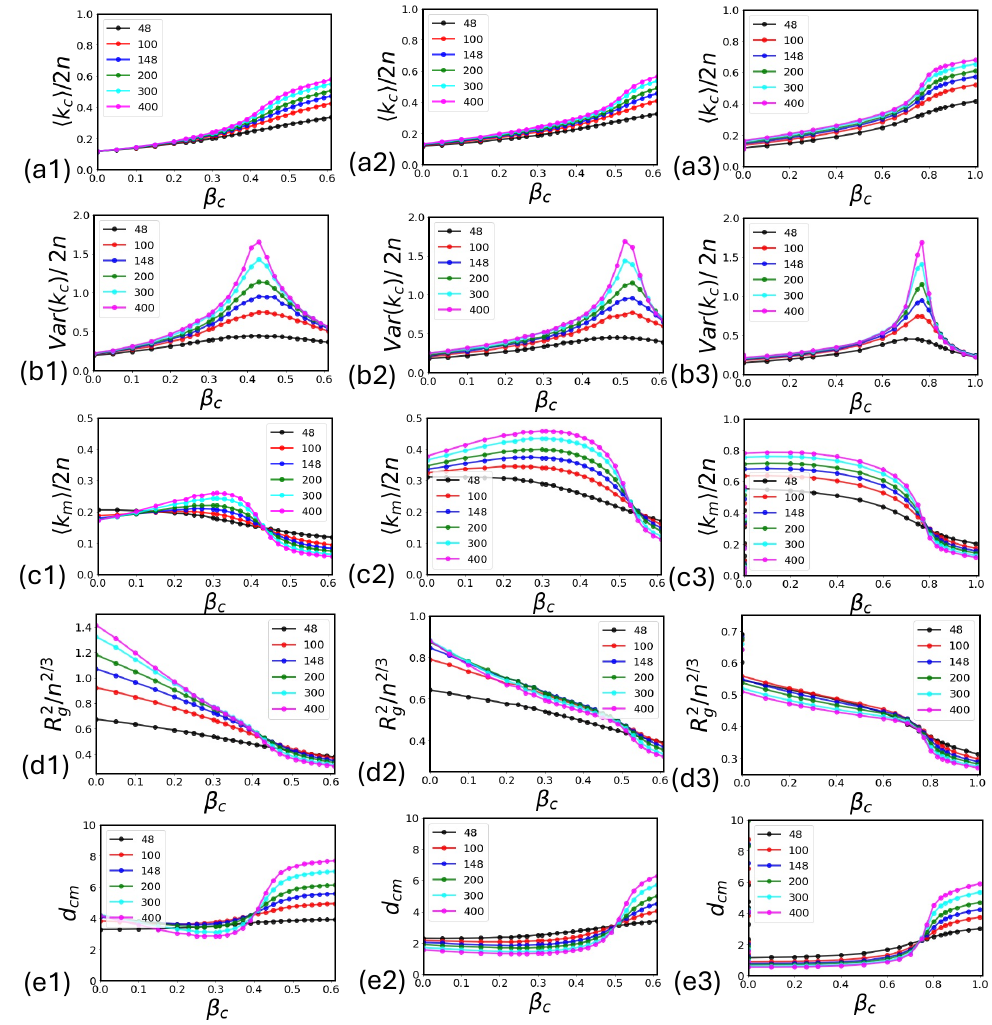}
    \caption{\textbf{(a1-a3) The average number of self-contacts scaled by 
    the system's size, $\langle k_c \rangle / 2n$, as a function of $\beta_c$.
    (b1-b3) Variance of $k_c$ scaled by the system's size, 
    $Var(k_m)/2n$, as a function of $\beta_c$. 
    (c1-c3)
    The average number of mutual contacts scaled by the system's size, 
    $\langle k_m \rangle / 2n$, as a function of $\beta_m$.       
    (d1-d3) $\beta_m$ dependence of the mean-squared radius 
    of gyration of one component, $R_g^2$, scaled by $n^{2/3}$.
    (e1-e3) Corresponding plots of the distance between the 
    centre of mass of the two polygons, $d_{cm}$, as a function of $\beta_m$.
    Different symbols refer to different system sizes (see legend).  The 
    three columns $(1,2,3)$ correspond to simulations performed at 
    the fixed values of $\beta_m=0.40,0.50,0.75$, respectively.}
    }
\label{fig:M_MC_thermo}
\end{figure}

A further indication of an SE-M phase boundary is seen in the behaviour 
of the distance between the centres of mass of the two components, 
$d_{cm}$, as a function of $\beta_m$ (these data are plotted in 
the bottom row panels in figure~\ref{fig:SE_ME_thermo}).
In the SE phase, for $\beta_c=0$, the hard-core repulsion between
vertices in the two polygons induces an entropic repulsive force 
between the polygons, segregating them when $\beta_m<\beta_m^*(0)$.
Here the natural length scale is $n^\nu$, where $\nu$ is the metric
exponent of the self-avoiding walk.  This shows that $d_{cm} \sim n^\nu$
in the SE-phase.  In the M-phase the two-component polygons share
the same region as the attractive forces have overcome the entropic
repulsion due to hard core nearest neighour interactions.  In this 
case $d_{cm}$ should approach a constant, or grow slowly with $n$ at a rate
slower than $O(n^\nu)$.  These expectations are supported by
the data in the left-most panel in the bottom row of figure~\ref{fig:SE_ME_thermo},
and the corresponding data for $\beta_c=0.28$ seen in the other panel 
show consistent behaviour with $\beta_c=0$,  also indicating that 
the phase boundary between the SE- and M-phases
in figure~\ref{fig:sketch_pd} is a straight line.

To determine the approximate location of the critical point $\beta_m^*(\beta_c)$ 
the crossings between the curves $Var(k_m)/2n$ and
$d_{cm}$ were determined and plotted against $1/\sqrt{n}$ in 
figure~\ref{fig:betam_estimates_SE_ME} (assuming the mean field value $1/2$ 
for the finite size crossover exponent $\phi$ and plotting the intersections
between the variance for $n=400$ and for $n=48,100,148,200$ and $300$). 
The data plausibly extrapolate to the value seen in reference \cite{JOTW22}, namely 
$\beta_m^*(0)  \approx 0.31$ (the thick segment at $1/\sqrt{n}=0$ marks the estimate 
of the expanded-mixed phase transition $\beta_m^*(0) = 0.31 \pm 0.01$ reported in 
reference~\cite{JOTW22}).  A similar extrapolation is reported 
for the location of the crossings of the curve $d_{cm}(n=400)$ with 
the corresponding curves obtained for $n=48,100,148,200$ and 
$300$ in the last row of panels in figure~\ref{fig:betam_estimates_SE_ME}.  
The limiting estimate shown by the thick segment at $1/\sqrt{n}=0$ 
is again the expanded-mixed phase  transition point 
$\beta_m^*(0) = 0.31 \pm 0.01$ reported in reference~\cite{JOTW22}. 
Similarly to the SE-SC phase boundary, we note that: 
(i) the $n\to\infty$ extrapolation of the points of both measures is very 
close to $\beta_m^*(0)$, and (ii) the SE-M phase transition appears to be 
independent of the strength $\beta_c$ of the self attraction between 
vertices in the polygons, at least when $0 \leq \beta_c < \beta_c^0$.

\begin{figure}[t!]	
\includegraphics[width=1\textwidth]{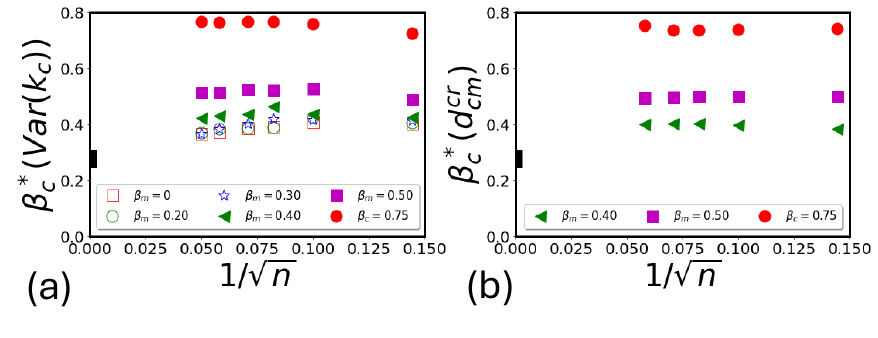}
	\caption{\textbf{Determining the locations of points on the critical curve
	$\beta_m^* (\beta_c)$ by extrapolating the locations of peaks in the $Var(k_c)$ 
	curves (denoted by $\beta_c^*(Var(k_c)$) and extrapolating the locations of 
	crossing points between $d_{cm}$ curves (denoted by 
	$\beta_c^*(d_{cm}^{cr})$) -- see figure \ref{fig:M_MC_thermo} panels 
	(b1), (b2) and (b3), and panels (e1), (e2) and (e3).
	(a) Comparison of the locations of the maxima of the 
	variance curves $Var(k_c)/2n$ for various values of $\beta_m$.  These
	data plausibly extrapolate to $\beta_c^0$ for data below the M-SE transition
	(this corresponds to the empty symbols), but to larger values above the
	M-SE transition (solid symbols).  
	(b) Similar to (a), but now for the separation between the centre-of-mass
	data $d_{cm}$ plotted for various values of $\beta_m$. The values of $\beta_m$ 
	are reported explicitly in the legend.  
	In both panels data have been plotted as a function of $1/\sqrt{n}$. The vertical 
	bars at $1/\sqrt{n}=0$ refers to the estimates of $\beta_c^0$ provided 
	in equation \Ref{eqn1}.}
	} 
\label{fig:betac_estimate_M_C}
\end{figure}

\subsection{The SC-M phase boundary}

Next, we consider the phase diagram in the region 
$\beta_c> \beta_c^0$ and $\beta_m > \beta_m^*(0)$.
We performed simulations crossing phase boundaries as shown in the 
right panel of figure~\ref{fig:simulations} by keeping fixed the parameter 
$\beta_c$ (mutual attractions) at values $\{0.4,0.5,0.75\}$ inside the 
SC phase while varying $\beta_m$ across SC-M phase boundaries. 
A similar approach was followed to examine the M-SC boundary by
varying $\beta_c$ while keeping $\beta_m$ fixed at the values
$\{0.4,0.5,0.75\}$, following the corresponding arrows in the left panel of
figure~\ref{fig:simulations}.

In figure~\ref{fig:CS_CM_thermo} $\beta_c=0.4$ in the first column, 
$\beta_c=0.5$ in the second column, and $\beta_c=0.75$ in the third 
column, while $\beta_m$ varies from $0$ to $1$.  In the first row of graphs 
we plot the average number of mutual contacts per unit length
$\langle k_m\rangle/2n$, in the second row the variance of $k_m$ per
unit length, $Var(k_m)/2n$, is plotted, and in the third row the separation 
$d_{cm}$ between the centres of mass of the component polygons is plotted.

Similar to the SE-M phase boundary (see section~\ref{SE-M}) the graphs
in figure~\ref{fig:CS_CM_thermo} shows the presence of finite-size
phase behaviour at a critical value of $\beta_m$.  The location of the
phase boundary can be estimated by again examining crossings
between the curves for the variance and $d_{cm}$ in 
figure~\ref{fig:CS_CM_thermo}. These are plotted in 
figure~\ref{fig:CS_CM_betac}  (the left panel is the data determined
from $Var(k_m)$ and the right panel corresponds to the results from
$d_{cm}$).  The data for $\beta_c=0,0.2,0.3$ are also included for
reference in the graphs (see figure~\ref{fig:betam_estimates_SE_ME})
-- these are the empty symbols.  Notice that these empty symbols
coincide within statistical uncertainty for different values of
$\beta_c=0,0.2,0.3$, but that when $\beta_c > \beta_c^0$ then the
data (the filled-in symbols) spread to larger values of $\beta_m$. This
trend is seen in both the left and right panels of figure~\ref{fig:CS_CM_thermo}.
In figure~\ref{fig:CS_CM_betac}, for $\beta_c=0.5$, the data 
extrapolate (purple triangles for $\beta_c=0.5$) to 
$\beta_m^*(0.5) \approx 0.5$, while for $\beta_c=0.75$ 
the data (red circles) seem to extrapolate to $\beta_m^*(0.75) \approx 0.75$. 
The case $\beta_c=0.4$ (yellow/green triangles) is more delicate 
since it is closer to the line of $\theta$-points along the SE-SC phase
boundary separating the SE-M and SC-M phase boundaries.  The 
results in the right panel apparently extrapolating to a value 
$\beta_m^*(0.4) \approx 0.4$ when $\beta_c=0.4$.

\subsection{The M-SC phase boundary}

In this section we consider data along lines crossing from the mixed into the
segregated-collapsed phase as illustrated in the left panel 
of figure~\ref{fig:simulations}.  In this case $\beta_m$ is fixed at 
one of $\{0.4,0.5,0.75\}$, while $\beta_c$ is increased in the interval
$[0,1]$.  In this case a phase boundary between the M and SC
phases should be crossed.  Our simulations tracked the observables 
plotted in figure~\ref{fig:M_MC_thermo}.  The left column of panels 
corresponds to $\beta_m=0.4$, the middle column to $\beta_m=0.5$, 
and the right column to $\beta_m=0.75$.  In this case the data are plotted 
as a function of $\beta_c\in [0,0.75]$.  

The following results are observed: 
\begin{description}
\item[Density of contacts $k_c$:] The density of nearest neighbour contacts 
$\langle k_c\rangle /2n$ is plotted in the top row of panels in figure~\ref{fig:M_MC_thermo}.  
    Here, the $\beta_c$ dependencies are similar to what was observed as we crossed 
    the SE-SC boundary in the top row of panels in figure~\ref{fig:SE_SC_thermo}.
\item[Variance of $k_c$:]  The variance develops a peak consistent with the
    location of the M-SC phase transition, and this peak grows with $n$ similarly
    to the peaks seen in the second row of panels in figure~\ref{fig:SE_SC_thermo}.  
    Moreover, the data show that $k_c$ is a sensitive indicator for detecting the 
    transition across the M-SC phase boundary.
\item[Density of mutual contacts $k_m$:]  The $\beta_c$ dependence of 
    $\langle k_m\rangle/2n$ has a more interesting profile than observed across 
    the SE-SC boundary in the third row of panels in figure~\ref{fig:CS_CM_thermo}.
    The curves have non-monotonic behaviour in the mixed phase with increasing
    $\beta_c$, first increasing for small $\beta_c$, but then decreasing as the
    M-SC phase boundary is approached and then crossed.  This behaviour
    is due to competition between self- and mutual contact interactions.  As $\beta_c$
    increases in the mixed phase, the component polygons contract, creating more
    mutual and self-contacts.  In the SC phase more self-contacts are formed to the
    exclusion of mutual contacts.    We also notice that the curves intersect at a 
    particular point insensitive to system size $n$.  This can be used to determine the
    location of the M-SC phase boundary.
\item[Radius of gyration:] Unlike the SE-SC boundary the behaviour of $R_g$ is 
    less clear.  It generally decreases with increasing $\beta_c$, as expected.  
    However, the scaling seems to have significant finite-size corrections.  
    For instance, at $\beta_m =0.4$ it does not scale proportionally to 
    $n^{2/3}$.  This suggests that the M phase is not a compact collapsed 
    regime.   As $\beta_c$ increases, crossing the M-SC phase boundary, 
    the expected $n^{2/3}$ behaviour appears to emerge.  In the case 
    of larger values of $\beta_m$ the curves in the middle and
    right panels collapse over the entire range of $\beta_c$ at larger 
    values of $n$, suggesting that the $n^{2/3}$ scaling emerges and 
    the phase has characteristics of a collapsed phase.
\item[Distance between the two centres of mass:] These are plotted in the bottom 
    row of panels in figure~\ref{fig:M_MC_thermo}.  The corresponding curves
    show non-linear behaviour, increasing as the M-SC phase boundary
    is crossed.  In the M phase the polygons are interpenetrating each other,
    resulting in a smaller separation $d_{cm}$.  Crossing into the SC phase
    increases the separation, as one would expect.  The non-linear behaviour of
    $d_{cm}$ with increasing $\beta_c$ in the M phase is more difficult to 
    explain, but there is a competition between mutual and self-contacts.  
    Increasing $\beta_c$ from a small value in the M phase would
    contract the component polygons, also increasing the mutual contacts.  However,
    as $\beta_c$ increases, self-contacts are formed at the cost of mutual contacts,
    tending to separate the polygons more as the critical M-SC phase boundary
    is approached.  
\end{description}

Estimates of the location of the M-SC phase boundary, based on 
peaks in the $Var(k_c)$ curves and the crossings between $d_{cm}$ curves are 
plotted in figure~\ref{fig:betac_estimate_M_C}.  These points are again plotted against
$1/\sqrt{n}$, and the vertical bar on the $y$-axis is the confidence interval
of estimate of $\beta_m^*$ in reference~\cite{JOTW22} 
when $\beta_c=0$.

\begin{figure}[t!]	
\includegraphics[width=1\textwidth,height=0.5\textheight]{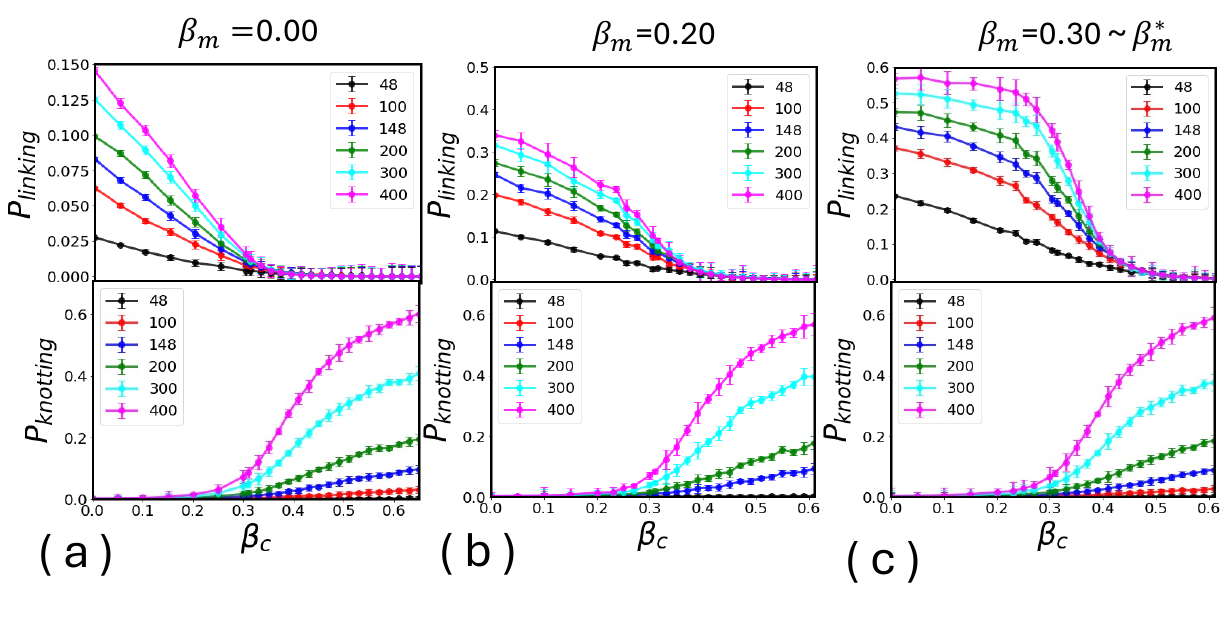}
	\caption{\textbf{Topological entanglement as the SE-SC phase 
	boundary is crossed by varying $\beta_c$: Linking probability 
	(top row) and knotting probability (bottom row) as a function 
	of the self-attraction parameter $\beta_c$ for three fixed values 
	of the mutual interaction parameter $\beta_m$, (	a) $0.00$, 
	(b) $0.20$ and (c) $0.30$. Note that $\beta_m=0.30$ is very 
	close to the estimate of the transition point $\beta_m^*$ 
	when $\beta_c=0$.}
	} 
\label{fig:SE_SC_topo}
\end{figure}

\begin{figure}[t!]	
\includegraphics[width=1\textwidth,height=0.5\textheight]{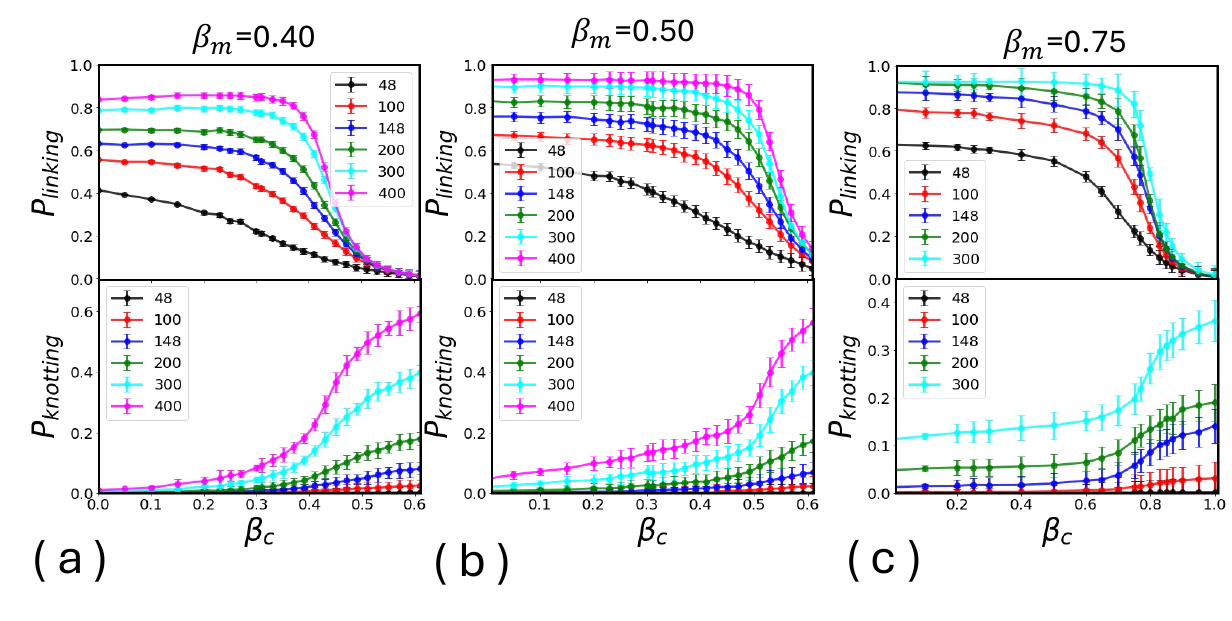}
	\caption{\textbf{Topological entanglement as the M-SC phase boundary 
	is crossed by varying $\beta_c$: Linking probability (top row) and 
	knotting probability (bottom row) as a function of the self-attraction 
	parameter $\beta_c$ for three fixed values of the mutual interaction 
	parameter $\beta_m$, (a) $0.40$, (b) $0.50$ and (c) $0.75$. All values 
	of $\beta_m$ correspond to the system well inside the mixed phase.}
	} 
\label{fig:ME_SC_topo}
\end{figure}

\section{Topological entanglement}
\label{sec:topology}

In the previous sections we have seen that the two polygons have
distinct metric and thermodynamic properties in the various thermodynamic
phases in figure \ref{fig:sketch_pd}.  For example, in the mixed phase, the
model is characterised by the mixing of the polygons, increasing mutual
contacts.  Characteristic in this phase is that the pair of polygons is
reduced in size with an increased local density of monomers inside the
convex hull of the pair.  This increased density should also increase
entanglements between strands of the two polygons, increasing the
probability that they are catenated into a non-trivial topological link.

In contrast, steric repulsion between monomers in the segregated phases
should make linking of the two components less likely.  In the 
segregated-expanded phase, the two components are separated from 
each other, and this should also reduce the probability that
each of the components is knotted (in addition to the
two components being unlikely to form a non-trivial link).  In the
segregated and collapsed phase, the two components remain separated,
and so again are unlikely to form a non-trivial link.  However, each 
component is now compacted and more likely to be knotted.

Thus, the (statistical) topological properties, as measured by the knotting 
of the components or the linking between components, change as the 
two rings cross phase boundaries.  In any given phase, the topological 
properties may vary continuously, but one expects sudden
changes across phase boundaries.  To explore these expectations, we
analysed the linking and knotting in the model as $\beta_m$ changes
while $\beta_c$ is kept fixed, or vice versa.

Linking between the $A$- and $B$-blocks is computed via the 
two-variable Alexander polynomial $\Delta(t,s)$ \cite{torres1953alexander}
while knot detection is based on the computation of the 
Dowker code \cite{dowker1983classification}.

Since their computation would be prohibitively costly when the 
number of crossings $n_c$ after a planar projection is very large, 
we first simplify the geometrical entanglement of the system while 
keeping its topology unaltered \cite{van1990knot}. 
This is achieved by shrinking stochastically the polygons with the BFACF 
algorithm, simulated at a very small step fugacity $K$ 
\cite{1981bergrandom,de1983new,marco2014knotted}. The BFACF moves 
have the property of preserving the topology and, if $K$ is sufficiently 
small, the system will often eventually shrink to its minimal length compatible 
with its topology. Even if the minimal length is not always reached, the shrunk 
configuration, once projected onto a plane, displays, most of the time, a 
sufficiently small number of crossings (well below $50$) to dramatically 
reduce the time required to compute the topological 
invariants \cite{van1990knot,millett2013identifying}.

\begin{figure}[t!]	
\includegraphics[width=1\textwidth,height=0.5\textheight]{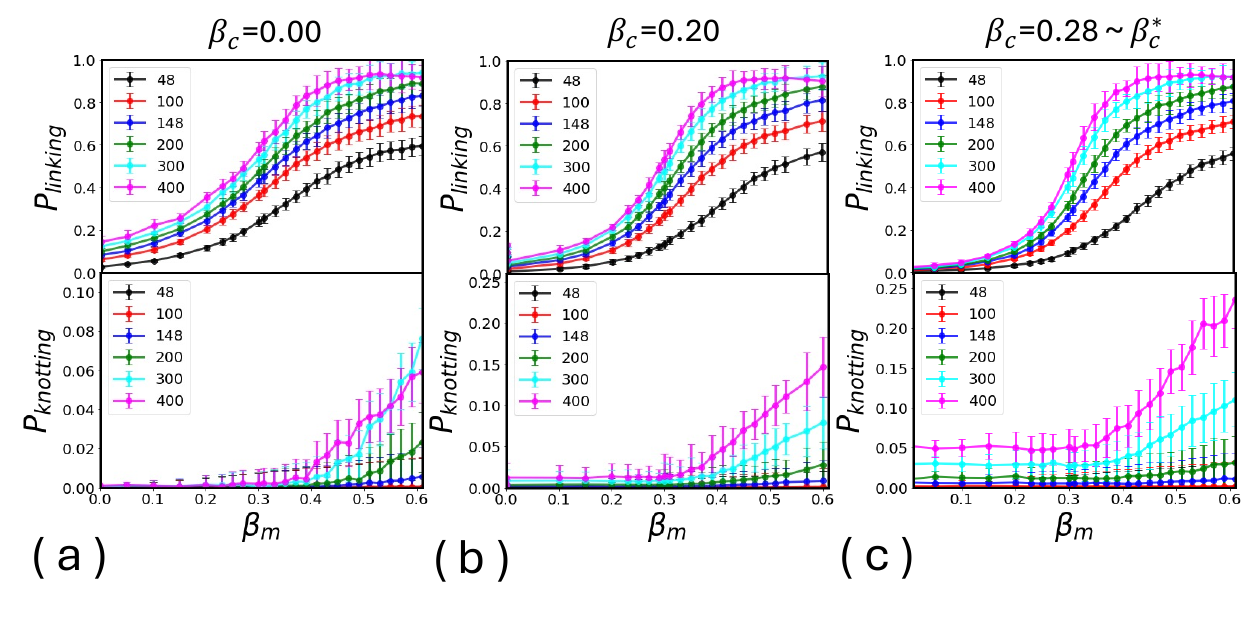}
\caption{\textbf{Topological entanglement as the SE-M phase boundary is 
crossed by varying $\beta_m$: Linking probability (top row) and knotting probability 
(bottom row) as a function of the mutual attraction parameter $\beta_m$ 
for three fixed values of the self attraction parameter $\beta_c$, (a) $0.00$,
(b) $0.20$ and (c) $0.28$. Note that $\beta_c=0.28$ 
is very close to the estimate of the $\theta$-point $\beta_c^0$  
(where the collapse transitionoccurs) when $\beta_m=0$.}
} 
\label{fig:SE_ME_topo}
\end{figure}

\subsection{Fixing $\beta_m$ and varying $\beta_c$}

In this section we fix $\beta_m$ while increasing $\beta_c$.  
The values of $\beta_m$ are chosen as illustrated in the
left panel of figure \ref{fig:simulations} and the trajectories in the phase
diagram crossing the SE-SC phase boundary when $\beta_m\in\{0,0.2,0.3\}$
and the M-SC phase boundary when $\beta_m\in\{0.4,0.5,0.75\}$.

In figure~\ref{fig:SE_SC_topo} the linking and knotting properties of the model are 
explored for $\beta_m\in\{0,0.2,0.3\}$ so that the SE-SC phase boundary is crossed
when $\beta_c$ increases beyond the critical point $\beta_c^0 \approx 0.28$.
In the SE phase (the top row of panels in the figure for $\beta_c < \beta_c^0$) 
the components are linked with positive probability, increasing as the 
length of the polygons increases and as $\beta_m$ approaches from 
below the transition value $\beta_m^*(\beta_c)$.  Increasing $\beta_c$ reduces the probability
of linking to zero in the SC phase.  This behaviour is expected, as the two 
components are segregated and compact in this phase, excluding each other
and so making linking unlikely.  In contrast, in the SE phase, the two polygons,
although segregated, are also expanded, allowing strands of one to enter
the other and to become entangled, giving rise to a positive probability of 
topological linking between the two components.

Knotting of the component polygons should increase when the SE-SC 
phase boundary is crossed, and this is seen in the bottom row of 
panels in figure~\ref{fig:SE_SC_topo}.  The knotting probability 
increases from a small value to significant values in the SC phase, 
consistent with previous studies showing low knotting probabilities 
of expanded polygons in the cubic lattice, which increase rapidly as 
the $\theta$-point is crossed into a phase of 
collapsed polygons \cite{tesi1994knotting,mansfield2007development}.

In figure~\ref{fig:ME_SC_topo} we show similar data, but when the M-SC 
phase boundary is crossed by fixing $\beta_m$ and increasing $\beta_c\in(0,1.0)$
(where $\beta_m\in\{0.4,0.5,0.75\}$.  In the M phase, the two 
polygons are mixed, and so we expect enhanced probabilities of linking.
This is clearly evident in the top row of panels of figure~\ref{fig:ME_SC_topo} 
and can be seen by comparing the data to the corresponding panels in 
figure~\ref{fig:SE_SC_topo}.  In addition, as the M-SC phase boundary 
is approached, the linking probabilities in these figures decrease quickly 
into the SC phase.  The results on knotting shown in the bottom row of 
panels in figure~\ref{fig:ME_SC_topo} are consistent with what is seen 
for the SE-SC transition in figure~\ref{fig:SE_SC_topo}.  The knotting
probability is low and not much increased in the M phase, compared
to the SE phase, but increases as the M-SC phase boundary is crossed 
in the collapsed phase, consistent with our observations above.

\begin{figure}[t!]	
\includegraphics[width=1\textwidth,height=0.5\textheight]{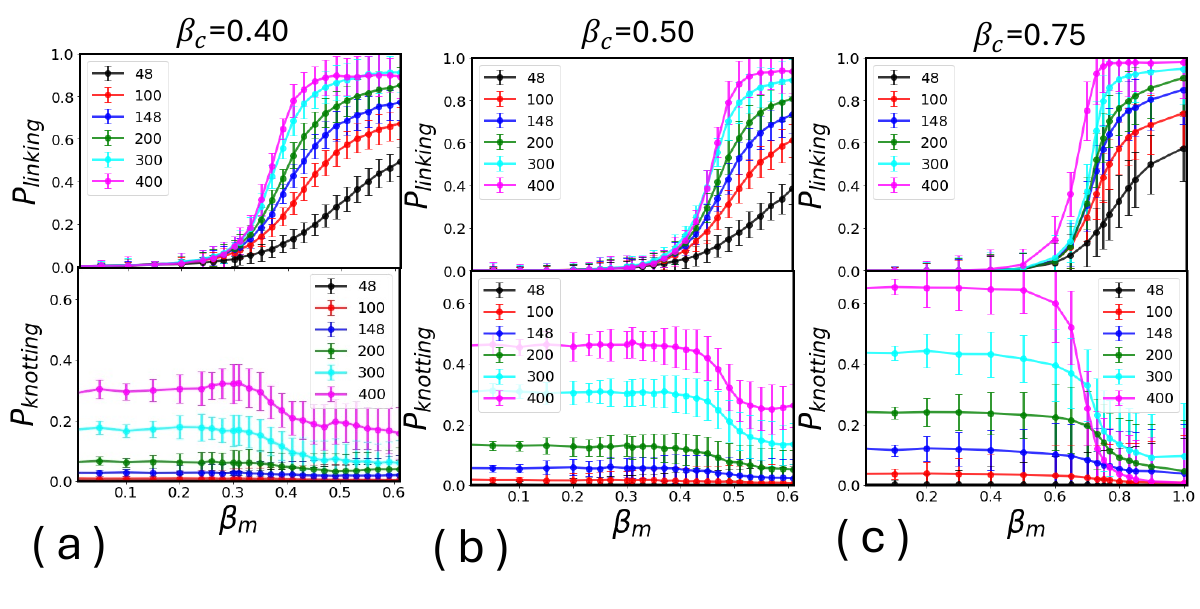}
\caption{\textbf{Topological entanglement as the SC-M phase boundary is crossed by 
varying $\beta_m$: Linking probability (top row) and knotting probability 
(bottom row) as a function of the mutual attraction parameter $\beta_m$ 
for three fixed values of the self attraction parameter $\beta_c$, (a) $0.40$, 
(b) $0.50$ and (c) $0.758$; all these values correspond to the system being 
well inside the collapse phase, at least for $\beta_m=0$.}
} 
\label{fig:SC_ME_topo}
\end{figure}

\subsection{Fixing $\beta_c$ and varying $\beta_m$}

By fixing $\beta_c$ and then increasing $\beta_m\in (0,0.75)$ one
can cross the SE-M phase boundary when $\beta_c$ is small enough,
or the SC-M phase boundary for $\beta_c$ large enough (this is
evident in the right panel of figure \ref{fig:simulations}).

The probability of knotting and linking when the model crosses the 
SE-M phase boundary is shown in the top and bottom rows of panels in 
figure~\ref{fig:SE_ME_topo}, respectively.  Linking probabilities increase 
dramatically from near zero in the SE phase as the SE-M phase 
boundary is crossed. Knotting probability, on the other hand, increases mildly 
across this phase boundary.  It is near zero in the SE
phase, and shows a small, though significant, increase in the M phase.

For larger $\beta_c$ (i.e., above the $\theta$-point), the transition is from the 
SC phase to the M phase, and the $\beta_m$ dependence of the linking 
and knotting probability is reported in figure~\ref{fig:SC_ME_topo}.  More specifically, 
the top row of panels shows that the linking probability (near zero in the SC phase)
increases to very large values in the M phase with increasing $\beta_m$.
Knotting probabilities, on the other hand, (see bottom row of panels) decrease 
as the SC-M phase boundary is crossed, from relatively large values in the SC 
phase, to smaller (but still significant) values in the M phase.

\section{Discussion}
\label{sec:discussion}

In this paper we have constructed and analysed a lattice model of two 
ring polymers,  close together in space.  Each of the rings can collapse 
to a compact ball and, in addition,  the rings can interpenetrate 
to form a mixed phase.  The model contains two parameters: $\beta_m$ 
controlling the extent of mixing, and a collapse parameter $\beta_c$.  We 
have investigated the form of the phase diagram in the 
$(\beta_c,\beta_m)$-plane.  We find evidence for three phases:  
A segregated-expanded (SE) phase, a segregated-compact (SC) phase, 
and a mixed (M) phase.

In section 2, we describe the model in more detail and make comments 
about the phase diagram.  We provide some rigorous results in section 3, 
where we prove existence of a phase boundary between the 
segregated phases and the mixed phase.  The two parts of this boundary 
are the SE-M and SC-M boundaries in figure \ref{fig:sketch_pd}.   In 
addition we argue that the phase boundary between the SE and SC phases 
is a line at constant $\beta_c = \beta_c^0$.

In section 4, we give a brief description of the Monte Carlo method 
we use to investigate the phases in the phase diagram.  We performed 
extensive simulations using a Multiple Markov Chain implementation \cite{geyer1995annealing,tesi1996monte,tesi1996interacting}
of the Metropolis Algorithm \cite{metropolis1953monte} on lattice polygons
using pivot elementary moves \cite{madras1990monte} 
and local Verdier-Stockmayer style moves \cite{verdier1962monte}.  

We discuss our results in sections 5 and 6.  In section 5.1 we focus on 
thermodynamic properties of the model when the SE-SC phase boundary 
is crossed.  In section 5.2 we give similar results when we cross the 
SE-M phase boundary.  These transitions appear to be continuous.   
We examine the SC-M phase boundary in sections 5.3 and 5.4.  

In section 6, we turn our attention to topological properties 
(knotting and linking) as the various phase boundaries are crossed.

\section*{Acknowledgement}
 EJJvR acknowledges financial support 
from NSERC Canada) in the form of Discovery Grant RGPIN-2019-06303. 
Maria Carla Tesi is a member of  GNAMPA (Gruppo Nazionale per l’ Analisi 
Matematica, la Probabilità  e le loro Applicazioni)  of INdAM (Istituto Nazionale di Alta Matematica).

\bibliographystyle{unsrt}
\bibliography{References.bib}

\end{document}